\documentclass[a4paper,11pt]{article}
\usepackage[numbers,sort&compress]{natbib}
\usepackage[all]{xy}
\usepackage{amsmath,amsfonts,amssymb,amsthm,epsfig,amscd,comment,latexsym,psfrag}
\usepackage{CJK}
\usepackage{mathrsfs}
\usepackage{nicefrac,xspace,tikz}
\usepackage{arydshln}
\usetikzlibrary{arrows}
\usepackage{graphicx}
\usepackage{caption,subcaption}
\usepackage{xcolor,float}
\usepackage[title]{appendix}
\usepackage{graphicx}
\usepackage{caption}
\makeatletter

\newcommand{\Rmnum}[1]{\expandafter\@slowromancap\romannumeral #1@}
\makeatother

\def\footnoterule{\kern 1mm \hrule width 7cm \kern 2.2mm}%

\def\dsum{\displaystyle\sum}

\def\tr{\mathrm{tr}}

\allowdisplaybreaks
\begin{document}

\begin{center}
{\Large\bf $W$-representations of two-matrix models with infinite set of variables}\vskip .2in
{\large
Lu-Yao Wang$^{a}$,\footnote{wangly100@outlook.com}
Yu-Sen Zhu$^{a}$,\footnote{ zhuyusen@cnu.edu.cn}
Ying Chen$^{b}$,\footnote{chenying$\_$math@jsnu.edu.cn}
Bei Kang$^{c}$\footnote{Corresponding author:kangbei@ncwu.edu.cn} } \vskip .2in
$^a${\small{ School of Mathematical Sciences, Capital Normal University,
Beijing 100048, China}} \\
$^b${\small{School of Mathematics and Statistics, Jiangsu Normal University, Xuzhou 221116, Jiangsu, China }} \\
$^c${\small{ School of Mathematics and Statistics, North China University of Water Resources and Electric Power,
Zhengzhou 450046, Henan, China}}\\

\begin{abstract}

The Hermitian, complex and fermionic two-matrix models with infinite set of variables are constructed. We show
that these two-matrix models can be realized by the $W$-representations. In terms of the $W$-representations,
we derive the compact expressions of correlators for these two-matrix models.

\end{abstract}
\end{center}


\section{Introduction}
Matrix models have been developed to solve non-perturbative two-dimensional gravity and provide a rich set of approaches
to physical systems. For two-matrix model, there is the interaction between the two matrices. Hence it possesses a richer
mathematical structure than single matrix models, and thus produces more applications in physics and mathematics.
The two-matrix models have been studied as an important solvable example of statistical mechanical systems, i.e., Ising spins \cite{Chadha1981,David1985,Kazakov1986}.
For fermionic two-matrix model, the complete sets of loop equations can be derived \cite{Semenoff1996}.
The Ward identities in Kontsevich-like one-matrix models are used to relate the degree of potential in Kontsevich-like two-matrix
model to the $W$-constraints \cite{Marshakov9201010}. The spectral curves, loop equations and topological expansion for Hermitian
two-matrix models were presented in Refs.\cite{Kazakov2002,Eynard2005,Bergere2012}.

For $W$-representation of matrix model, it realizes partition function by acting on elementary functions with exponents
of the given $W$-operator \cite{Shakirov2009}. Since $W$-representation plays an important role in understanding the structures
of matrix models, much interest has been attributed to this direction. A variety of matrix models have been realized by
$W$-representations and their correlators can be exactly calculated. Recently the (super) partition function hierarchies with
$W$-representations were constructed \cite{WRLF2206,WRLF2208}. Some well known superintegrable matrix models were contained
in these superintegrable hierarchies. In addition, the progress of $W$-representation has been made on tensor models \cite{Itoyama1910,Kang2104,lywang2120,Kang2301} and
super-eigenvalue models \cite{Chen,wang2020}.

Recently, the two-matrix models with multi-set of variables were proposed \cite{Alexandrov2022,MMMP,MMMPZ,wly2023}, which are
the superintegrable matrix models. Their $W$-representations and character expansions were well investigated.
In this paper, we'll construct the new two-matrix models with infinite set of variables and derive their $W$-representations.

\section{$W$-representation of new Hermitian two-matrix model}
Let us construct the Hermitian two-matrix model
\begin{eqnarray}\label{partition-H}
Z_{2H}&=&\int dAdB \exp(-\frac{1}{2} \tr A^2-\frac{1}{2}\tr
B^2+\sum_{k=0}^{\infty}t_k \tr A^k
+\sum_{k=0}^{\infty}g_k \tr B^k
\nonumber\\&&
+\sum_{l=1}^{\infty}\sum_{k_1,\cdots k_{2l}=1}^{\infty}
t_{k_{1,\cdots ,2l}}
\tr A^{k_1}B^{k_2}A^{k_3}B^{k_4}\cdots A^{k_{2l-1}}B^{k_{2l}}),
\end{eqnarray}
where~$A$ and $B$ are~$N\times N$ matrices.
When $B=0$ in (\ref{partition-H}), it reduces to the well known
Gaussian Hermitian matrix model.

By requiring the invariance of the integral (\ref{partition-H}) under the infinitesimal transformation
$A\rightarrow A+\epsilon A^n\ (n\geq0)$ or $B\rightarrow B+\epsilon B^n\ (n\geq0)$,
we obtain the Virasoro constraints
\begin{eqnarray}\label{vira}
\mathcal{L}_{n-1} Z_{2H}=0,
\end{eqnarray}
where the constraint operators are given by
\begin{eqnarray}\label{viraconst}
\mathcal{L}_{n-1}&=&\sum_{n=0}^{\infty}2N\frac{\partial}{\partial t_{n-1}}
+\sum_{n=0}^{\infty}\sum_{s=1}^{n-1}\frac{\partial^2}{\partial t_{s}\partial t_{n-1-s}}+\delta_{n,1}N^2
+\sum_{n=0}^{\infty}\sum_{k=0}^{\infty}kt_k\frac{\partial}{\partial t_{n+k-1}}
+\sum_{k_2=1}^{\infty}t_{1,k_2}\frac{\partial}{\partial g_{k_2}}
\nonumber\\
&&
+\sum_{l=1}^{\infty}\sum_{k_1,\cdots,k_{2l}=1}^{\infty}
 t_{k_{1,\cdots ,2l}}
[
\sum_{a=1}^{l}\sum_{n=0}^{\infty}k_{2a-1}
\frac{\partial}{\partial t_{k_1,\cdots,n+k_{2a-1},k_{2a},\cdots,k_{2l}}}
+\delta_{k_1,1}\frac{\partial}{\partial t_{k_3,\cdots,k_{2l-1},k_{2l}+k_2}}
\nonumber\\
&&
+\sum_{a=2}^l\delta_{k_{2a-1},1}
\frac{\partial}{\partial t_{k_1,\cdots,k_{2a-2}+k_{2a},k_{2a+1},\cdots,k_{2l}}}
]
,
\end{eqnarray}
which obey the Virasoro algebra
\begin{eqnarray}\label{witt-h}
[\mathcal{L}_{n-1},\mathcal{L}_{m-1}]=(n-m)\mathcal{L}_{n+m-2}.
\end{eqnarray}

Let us now consider the following five infinitesimal transformations, respectively,

(i) $A\longrightarrow A+\epsilon\dsum_{n=0}^{\infty}(n+1)t_{n+1}A^n$,
(ii) $A\longrightarrow A+\epsilon\dsum_{n=1}^{\infty}(n+1)t_{1,n}B^n$,

(iii) $A\longrightarrow A+\epsilon \dsum_{r=1}^{\infty}\dsum_{n_1,\cdots n_{2r}=1}^{\infty}
\mathcal{N}_1 t_{n_1+1,n_2,\cdots,n_{2r}}
A^{n_1}B^{n_2}\cdots A^{n_{2r-1}}B^{n_{2r}}$,

(iv) $A\longrightarrow A+\epsilon \dsum_{r=1}^{\infty}\dsum_{n_1,\cdots n_{2r}=1}^{\infty}\mathcal{N}_2
t_{1,n_2,\cdots,n_{2r}} B^{n_2}A^{n_3}\cdots A^{n_{2r-1}}B^{n_{2r}}$,

(v) $B\longrightarrow B+\epsilon\dsum_{n=0}^{\infty}(n+1)g_{n+1}B^n$,

where $\mathcal{N}_1=n_1+1+n_2+\cdots+n_{2r}$ and $\mathcal{N}_2=1+n_2+\cdots+n_{2r}$.

From the invariance of the integral (\ref{partition-H}), it gives
\begin{eqnarray}\label{con-H1}
\hat{D}_iZ_{2H}=\hat{W}_iZ_{2H}, i=1,2,\cdots 5,
\end{eqnarray}
where the operators $\hat{W}_i$ are listed in (\ref{W-H}) and $\hat{D}_i$ are
\begin{eqnarray}\label{D-H}
\begin{array}{ll}
\hat{D}_1
=\dsum_{i=1}^{\infty}it_i\frac{\partial}{\partial t_i},
~~~~
\hat{D}_2
=\dsum_{n=1}^{\infty}(n+1)t_{1,n}\frac{\partial}{\partial t_{1,n}},
\\
\hat{D}_3
=\dsum_{r=1}^{\infty}\sum_{n_1,\cdots
n_{2r}=1}^{\infty}  \mathcal{N}_1t_{n_1+1,n_2,\cdots,n_{2r}}
\frac{\partial}{\partial t_{n_1+1,n_2,\cdots,n_{2r}}},
\\
\hat{D}_4
=\dsum_{r=1}^{\infty}\dsum_{n_1,\cdots
n_{2r}=1}^{\infty}\mathcal{N}_2t_{1,n_2,\cdots,n_{2r}}
\frac{\partial}{\partial t_{1,n_2,\cdots,n_{2r}}},
\ \
\hat{D}_5=\dsum_{i=1}^{\infty}ig_i\frac{\partial}{\partial g_i}.
\end{array}
\end{eqnarray}

In the following, we'll focus on the sum of (\ref{con-H1})
\begin{eqnarray}\label{constraint-H}
\hat DZ_{2H}=\hat WZ_{2H},
\end{eqnarray}
where $\hat D=\sum_{i=1}^{5}D_i$ and $\hat W=\sum_{i=1}^{5}W_i$.

Let us write the partition function (\ref{partition-H}) as the grading form
$Z_{2H}=\sum_{d=0}^{\infty}Z_{2H}^{(d)}$
and
\begin{eqnarray}
Z_{2H}^{(d)}
&=&e^{N(t_0+g_0)}\sum_{l=0}^{\infty}\frac{1}{l!}\sum_{\substack{l_1+l_2+l_3=l\\
\rho_1+\rho_2+\rho_3=d}}
\langle\prod_{i=1}^{l_1}\tr A^{k_i}\prod_{j=1}^{l_2}\tr
B^{r_j}\prod_{n=1}^{l_3}\tr A^{S_{n,1}}B^{S_{n,2}}\cdots
A^{S_{n,2p_n-1}}B^{S_{n,2p_n}}\rangle
\nonumber\\
&&
\cdot
\prod_{i=1}^{l_1}t_{k_i}\prod_{j=1}^{l_2}g_{r_j}\prod_{n=1}^{l_3}
t_{S_{n,1},\cdots,S_{n,2p_n}}
\cdot\int dAdB \exp(-\frac{1}{2}\tr A^2-\frac{1}{2}\tr B^2),
\end{eqnarray}
where~$\rho_1=\sum_{i=1}^{l_1}k_i,~\rho_2=\sum_{i=1}^{l_2}r_i$,
$\rho_3=\sum_{i=1}^{l_3}(S_{i,1}+\cdots+S_{i,2p_i})$ correlators~$\langle\cdots\rangle $ are defined as
\begin{eqnarray}\label{corcdots}
\langle\cdots\rangle=\frac{\int dAdB \cdots\exp(-\frac{1}{2} \tr
A^2-\frac{1}{2}\tr B^2)}{\int dAdB \exp(-\frac{1}{2} \tr
A^2-\frac{1}{2}\tr B^2)}.
\end{eqnarray}

We denote the degrees of operators as
$deg(t_k)=deg(g_k)=k$,
$deg(\frac{\partial}{\partial t_k})=deg(\frac{\partial}{\partial g_k})=-k$,
$deg(\frac{\partial}{\partial
t_{k_1,k_2,\cdots,k_{2l-1},k_{2l}}})=-(k_1+\cdots+k_{2l})$.
Then it is easy to see that $deg(\hat{D})=0$ and~$deg(\hat{W})=2$.

Due to~the operators $\hat{D}$ and $\hat{D}-\hat{W}$ being invertible and
$\hat{D}e^{N(t_0+ g_0)}=0$, from (\ref{constraint-H}), we have
\begin{eqnarray}
\dsum_{s=1}^{\infty}Z_{2H}^{(s)}
=(\hat{D}-\hat{W})^{-1}\hat{W}e^{N(t_0+g_0)}
=\dsum_{k=1}^{\infty}(\hat{D}^{-1}\hat{W})^{k}e^{N(t_0+g_0)}.
\end{eqnarray}
Note that~$\hat{W}$ is an homogeneous operator with degree~$2$,
and~$\hat{D}f=deg(f)\cdot f$ for any homogeneous function~$f$.
We may give the $W$-representation of the Hermitian two-matrix model (\ref{partition-H})
\begin{eqnarray}\label{w-rep-H}
Z_{2H}=e^{\frac{1}{2}\hat{W}}e^{N(t_0+g_0)}.
\end{eqnarray}

Let us formally write the $(m+1)$-th power of the operator $\hat{W}$  as
\begin{eqnarray}\label{W-m+1}
\hat{W}^{(m+1)}&=&\sum_{l_1+l_2+l_3=1}^{2(m+1)}
\sum_{\rho_1+\rho_2+\rho_3=2(m+1)}
P_{(k_1,\cdots,k_{l_1});(r_1,\cdots,r_{l_2})}
^{(S_{1,1},\cdots,S_{1,2p_1};\cdots;S_{l_3,1},\cdots,S_{l_3,2p_{l_3}})}
t_{k_1}\cdots t_{k_{l_1}}g_{r_1}\cdots g_{r_{l_2}}
\nonumber\\&&\cdot
t_{S_{1,1},\cdots,S_{1,2p_1}}\cdots
t_{S_{l_3,1},\cdots,S_{l_3,2p_{l_3}}}
+\cdots.
\end{eqnarray}

By means of the $W$-representation of (\ref{partition-H}), we derive the compact expression of correlators
\begin{eqnarray}\label{correlator-H}
&&\langle\prod_{i=1}^{l_1} \tr A^{k_i}\prod_{j=1}^{l_2} \tr
B^{r_j}\prod_{n=1}^{l_3} \tr A^{S_{n,1}}B^{S_{n,2}}\cdots
A^{S_{n,2p_n-1}}B^{S_{n,2p_n}}\rangle
\nonumber\\
&&=\frac{l_1!l_2!l_3!\dsum_{\rho_1+\rho_2+\rho_3=1}^{2(m+1)}\dsum_{\sigma}
P_{(\sigma(k_1),\cdots,\sigma(k_{l_1}));(\sigma(r_1),\cdots,\sigma(r_{l_2}))}
^{(\sigma(S_{1,1}),\cdots,\sigma(S_{1,2p_1});\cdots;\sigma(S_{l_3,1}),\cdots,
\sigma(S_{l_3,2p_{l_3}}))}}
{2^{m+1}(m+1)!\lambda_{(k_1,\cdots,k_{l_1})}\lambda_{(r_1,\cdots,r_{l_2})}
\lambda_{(S_{1,1},\cdots,S_{1,2p_1};\cdots;S_{l_3,1},\cdots,S_{l_3,2p_{l_3}})}},
\end{eqnarray}
where $(\sigma(k_1),\cdots,\sigma(k_{l_1}))$ denotes all distinct permutations of
$(k_{l_1},\cdots,k_{l_1})$,
and $\lambda_{(k_1,\cdots,k_{l_1})}$
is the number of distinct permutations $(k_1,\cdots,k_{l_1})$.

For example, let us consider the cases
\begin{eqnarray}
\hat{W}&=&t_1^2N+2t_2N^2+g_1^2N+2g_2N^2+\cdots ,\nonumber\\
\hat{W}^2&=&8t_1^2t_2N+24t_1t_3N^2+12t_4N^3+8t_2^2N^2+8t_1t_{1,2}N^2+8t_1g_1t_{1,1}N+8g_1t_{2,1}N^2
\nonumber\\
&&+4t_{1,1}^2N^2+8t_{2,2}N^3+8g_1^2g_2N+24g_1g_3N^2+12g_4N^3+8g_2^2N^2+\cdots.
\end{eqnarray}
We may give some correlators in (\ref{correlator-H}) as follows
\begin{eqnarray}
\begin{array}{lll}
\langle \tr A \tr A\rangle=\langle \tr B \tr B\rangle=N,&
~\langle \tr A^2 \rangle=\langle \tr B^2 \rangle=N^2,\\
\langle \tr A\tr B\tr A^2\rangle=2N+N^3,  &
 ~\langle \tr A \tr A^3\rangle=\langle\tr B\tr B^3\rangle= 3N^2, \\
\langle\tr A^4\rangle=\langle\tr B^4\rangle=3N^2, &
 ~\langle\tr A^2B^2\rangle=N^2, \\
\langle\tr AB\tr AB \rangle=N^2, &
  ~\langle\tr A\tr B\tr AB \rangle=N,\\
\langle \tr A \tr AB^2\rangle=\langle\tr A\tr A^2B\rangle=N^2,  &
 ~\langle \tr A^2 \tr A^2\rangle=\langle \tr B^2 \tr B^2\rangle=2N^2+N^4. &
\end{array}
\end{eqnarray}

\section{$W$-representation of complex two-matrix model}
Let us construct the complex two-matrix model
\begin{eqnarray}\label{partition-C2}
Z_{2C}&=&\int d^2M_1d^2M_2 \exp[-\mu \tr M_1M_1^{\dag}-\mu\tr
M_2M_2^{\dag}+\sum_{k=0}^{\infty}t_k \tr ( M_1M_1^{\dag})^k
+\sum_{k=0}^{\infty}g_k \tr ( M_2M_2^{\dag})^k
\nonumber\\
&&+\sum_{l=1}^{\infty}\sum_{k_1,\cdots k_{2l}=1}^{\infty}
t_{k_1,\cdots k_{2l}}\tr (M_1M_1^{\dag})^{k_1}(M_2M_2^{\dag})^{k_2}\cdots (M_1M_1^{\dag})^{k_{2l-1}}(
M_2M_2^{\dag})^{k_{2l}}],
\end{eqnarray}
where~$M_1$ and $M_2$ are $N\times N$ complex matrices.

By requiring the invariance of the integral (\ref{partition-C2}) under the infinitesimal transformation
$M_1\longrightarrow M_1+\epsilon(M_{1}M_{1}^{\dag})^nM_1\ (n\geq0)$
or $M_2\longrightarrow M_2+\epsilon(M_{2}M_{2}^{\dag})^nM_2~ (n\geq0)$,
it gives the Virasoro constraints
\begin{eqnarray}
\bar{\mathcal{L}}_n Z_{2C}=0,
\end{eqnarray}
where
\begin{eqnarray}\label{viraconst-2c}
\bar{\mathcal{L}}_n &=&\sum_{n=0}^{\infty}2N\frac{\partial}{\partial t_{n}}
+\sum_{n=0}^{\infty}\sum_{s=1}^{n-1}\frac{\partial^2}{\partial t_{s}\partial t_{n-s}}+\delta_{n,0}N^2
-\mu\sum_{n=0}^{\infty}\frac{\partial}{\partial t_{n+1}}
+\sum_{n=0}^{\infty}\sum_{k=0}^{\infty}kt_k\frac{\partial}{\partial t_{n+k}}
\nonumber\\
&&
+\sum_{n=0}^{\infty}\sum_{l=1}^{\infty}\sum_{k_1,\cdots,k_{2l}=1}^{\infty}
\sum_{a=1}^{l}k_{2a-1} t_{k_1,\cdots,k_{2l}}
\frac{\partial}{\partial t_{k_1,k_2,\cdots,n+k_{2a-1},k_{2a},\cdots,k_{2l}}}.
\end{eqnarray}

Similarly, the four constraints of (\ref{partition-C2}) can be derived from the invariance of the integral
under the following four infinitesimal transformations, respectively,

(i) $M_1\longrightarrow M_1+\epsilon\dsum_{n=0}^{\infty}(n+1)t_{n+1}(M_{1}M_{1}^{\dag})^nM_1$,

(ii) $M_1\longrightarrow M_1+\epsilon\dsum_{n,m=0}^{\infty}[(n+1)+(m+1)]t_{n+1,m+1}
      (M_{2}M_{2}^{\dag})^{m+1}(M_{1}M_{1}^{\dag})^{n}M_1$,

(iii) $M_1\longrightarrow M_1+\epsilon\dsum_{r=1}^{\infty}\dsum_{n_1,\cdots,n_{2r+1}=0}^{\infty}\bar{\mathcal{N}}
     t_{n_{2r+1},n_2+1,\cdots,n_{2r}+1}(M_{2}M_{2}^{\dag})^{n_2+1}(M_{1}M_{1}^{\dag})^{n_3+1}\cdots
     \\~~~~~~~~~~~~~~~~~~~~~
     \cdots(M_{2}M_{2}^{\dag})^{n_{2r}+1}(M_{1}M_{1}^{\dag})^{n_{2r+1}}M_1$,

(iv) $M_2\longrightarrow M_2+\epsilon\dsum_{m=0}^{\infty}(m+1)g_{m+1}(M_{2}M_{2}^{\dag})^mM_2$,

where $\bar{\mathcal{N}}=(n_2+1)+(n_3+1)+\cdots+(n_{2r+1}+1)$.
The sum of these constraints are
\begin{eqnarray}\label{constraint-2C-1}
\mu\bar{D}Z_{2C}=\bar{W}Z_{2C},
\end{eqnarray}
where $\bar{D}=\sum_{i=1}^4\bar{D}_i$, $\bar{W}=\sum_{i=1}^4\bar{W}_i$, the operators $\bar{D}_i$ are
\begin{eqnarray}\label{D-2C-1}
\begin{array}{ll}
\bar{D}_1
=\dsum_{n=0}^{\infty}(n+1)t_{n+1}\frac{\partial}{\partial t_{n+1}},
\qquad\qquad\quad
\bar{D}_2
=\dsum_{n,m=0}^{\infty} \bar T_1\frac{\partial}{\partial t_{n+1,m+1}},
\\
\bar{D}_2=\dsum_{n_1,\cdots,n_{2r+1}=0}^{\infty}\bar T_2\frac{\partial}{\partial t_{n_1+1,\cdots,n_{2r+1}+1}},
\ \
\bar{D}_4=\dsum_{m=0}^{\infty}(m+1)g_{m+1}\frac{\partial}{\partial g_{m+1}},
\end{array}
\end{eqnarray}
and the operators $\bar{W}_i$ are
\begin{small}
\begin{eqnarray}\label{W-2C-1}
\bar{W}_1&=&\sum_{n=0}^{\infty}(n+1)t_{n+1}[2N \frac{\partial}{\partial
t_{n}}(1-\delta_{n,0})
+\sum_{a=1}^{n-1}\frac{\partial}{\partial t_{a}}\frac{\partial}{\partial t_{n-a}}
+\sum_{k=0}^{\infty}kt_{k}\frac{\partial}{\partial
t_{n+k}}]
+N^2t_{1}
\nonumber\\&&
+\sum_{n=0}^{\infty}\sum_{l=1}^{\infty}\sum_{a=1}^{l}
\sum_{k_1,\cdots,k_{2l}=1}^{\infty}(n+1)t_{n+1}k_{2a-1} t_{k_1,\cdots,k_{2l}}
\frac{\partial}{\partial t_{k_1,\cdots,k_{2a-1}+n,k_{2a},\cdots,k_{2l}}},
\nonumber\\&&
\nonumber\\
\bar{W}_2&=&
\sum_{n,m=0}^{\infty}
\bar T_1
\{(N \frac{\partial}{\partial
t_{n,m+1}}
+\frac{\partial^2}{\partial g_{m+1}\partial t_{n}})(1-\delta_{n,0})
+\sum_{a=1}^{n-1}\frac{\partial^2}{\partial t_{a,m+1}\partial t_{n-a}}
+N\delta_{n,0}\frac{\partial}{\partial
g_{m+1}}
\nonumber\\&&
+\sum_{k=0}^{\infty}kt_{k}\frac{\partial}{\partial t_{n+k,m+1}}
+\sum_{l=1}^{\infty}\sum_{k_1,\cdots,k_{2l}=1}^{\infty} t_{k_1,\cdots,k_{2l}}
[k_1\frac{\partial}{\partial t_{n+k_1,\cdots,k_{2l-1},k_{2l}+m+1}}
\nonumber\\&&
+k_{2l-1}\sum_{a=2}^{l}\frac{\partial}{\partial
t_{k_1,\cdots,k_{2a-1}+n,k_{2a},\cdots,k_{2l}}}
+\sum_{a=1}^{l}\sum_{s=1}^{k_{2a-1}-1}\frac{\partial}{\partial
t_{k_1,\cdots,s,m+1,n+k_{2a-1}-s,k_{2a},\cdots,k_{2l}}}
]\}
\nonumber\\&&
\nonumber\\
\bar{W}_3&=&\sum_{n_1,\cdots,n_{2r+1}=0}^{\infty}
\bar T_2
\{[N\frac{\partial}{\partial g_{n_2+1}}\frac{\partial}{\partial
t_{n_3+n_{2r+1}+1,n_4+1,\cdots,n_{2r}+1}}
+N\frac{\partial}{\partial t_{n_3+1,\cdots,n_{2r}+n_2+2}}
\frac{\partial}{\partial t_{n_{2r+1}}}
\nonumber\\&&
+N\sum_{b=2}^{r-1}\frac{\partial}{\partial
t_{n_3+1,\cdots,n_{2b}+n_2+2}}\frac{\partial}{\partial
t_{n_{2b+1}+1,\cdots,n_{2r+1}}}
+\sum_{s=1}^{n_{2r+1}-1}
\frac{\partial}{\partial t_{s,n_2+1,\cdots,n_{2r}+1}}\frac{\partial}{\partial t_{n_{2r+1}-s}}
\nonumber\\&&
+\sum_{b=1}^{r-1}\sum_{s=1}^{n_{2b+1}}
\frac{\partial}{\partial t_{s,n_2+1,\cdots,n_{2b}+1}}
\frac{\partial}{\partial t_{\bar\xi_1, n_{2b+2}+1,\cdots,n_{2r}+1}}
+N\frac{\partial}{\partial t_{n_{2r+1},n_2+1,\cdots,n_{2r}+1}}]\cdot
\nonumber\\&&
\cdot(1-\delta_{n_{2r+1},0})
+
\delta_{n_{2r+1},0}
\frac{\partial}{\partial t_{n_3+1,\cdots,n_{2r-1}+1,n_{2r}+n_2+2}}
+\sum_{k=0}^{\infty}
kt_k\frac{\partial}{\partial t_{n_{2r+1}+k,n_2+1,\cdots,n_{2r}+1}}
\nonumber\\&&
+\sum_{l=1}^{\infty}\sum_{k_1,\cdots,k_{2l}=1}^{\infty}
t_{k_1,\cdots,k_{2l}}
[k_1\frac{\partial}{\partial t_{n_3+1,\cdots,\bar\xi_2, k_2,\cdots,\bar\xi_3}}
+\sum_{a=2}^l
(k_{2a-1}\frac{\partial}{\partial t_{k_1,\cdots,\bar\xi_4,k_{2a},\cdots,k_{2l}}}
 \nonumber\\&&
+\sum_{s=1}^{k_{2a-1}-1}\frac{\partial}{\partial t_{k_1,\cdots,k_{2a-2},s,n_2+1,\cdots,
  \bar\xi_5,k_{2a},\cdots,k_{2l}}})
]\},
\nonumber\\&&
\nonumber\\
\bar{W}_4&=&N^2g_{1}
+\sum_{m=0}^{\infty}(m+1)g_{m+1}[2N \frac{\partial}{\partial g_{m}}(1-\delta_{m,0})
+\sum_{a=1}^{m-1}\frac{\partial}{\partial g_{a}}
 \frac{\partial}{\partial g_{m-a}}+\sum_{k=0}^{\infty}kg_{k}\frac{\partial}{\partial g_{m+k}}]
\nonumber\\&&
+\sum_{m=0}^{\infty}\sum_{l=1}^{\infty}\sum_{a=1}^{l}
 \sum_{k_1,\cdots,k_{2l}=1}^{\infty}(m+1)g_{m+1}k_{2a}
 t_{k_1,\cdots,k_{2l}}
 \frac{\partial}{\partial t_{k_1,\cdots,k_{2a}+m,k_{2a+1},\cdots,k_{2l}}},
\end{eqnarray}
\end{small}
where
$\bar T_1=(n+m+2)t_{n+1,m+1}$,
$\bar T_2=\bar{\mathcal{N}}t_{n_{2r+1}+1,n_2+1,\cdots,n_{2r}+1}$
 and
~$\bar\xi_1=n_{2r+1}+n_{2b+1}+1-s$,
~$\bar\xi_2=n_{2r+1}+k_1$,
~$\bar\xi_3=k_{2l}+n_{2}+1$,
~$\bar\xi_4=(k_{2a-2}+n_2+1,n_3+1,\cdots,n_{2r}+1,n_{2r+1}+k_{2a-1})$,
~$\bar\xi_5=n_{2r+1}+k_{2a-1}-s$.

Similar to the case of the Hermitian two-matrix model (\ref{w-rep-H}),
the complex two-matrix model (\ref{partition-C2}) can be realized by the $W$-representation
\begin{eqnarray}\label{wrep-C2}
Z_{2C}=e^{\frac{1}{\mu}\bar{W}}e^{N(t_0+g_0)}.
\end{eqnarray}
There is also the compact expression of correlators
\begin{eqnarray}\label{correlator-C}
&&\langle\prod_{i=1}^{l_1} \tr (M_1M_1^{\dag})^{k_i}\prod_{j=1}^{l_2}
\tr (M_2M_2^{\dag})^{r_j}\prod_{n=1}^{l_3}
\tr (M_1M_1^{\dag})^{S_{n,1}}(M_2M_2^{\dag})^{S_{n,2}}\cdots
(M_2M_2^{\dag})^{S_{n,2p_n}}\rangle
\nonumber\\
&&
=\frac{l_1!l_2!l_3!\dsum_{\rho_1+\rho_2+\rho_3=1}^{m+1}\sum_{\sigma}
      \bar P_{(\sigma(k_1),\cdots,\sigma(k_{l_1}));(\sigma(r_1),\cdots,\sigma(r_{l_2}))}
            ^{(\sigma(S_{1,1}),\cdots,\sigma(S_{1,2_{p_1}});\cdots;\sigma(S_{l_3,1}),
            \cdots,\sigma(S_{l_3,2p_{l_3}}))}}
{\mu^{m+1}(m+1)!\lambda_{(k_1,\cdots,k_{l_1})}\lambda_{(r_1,\cdots,r_{l_2})}
 \lambda_{(S_{1,1},\cdots,S_{1,2p_1};\cdots;S_{l_3,1},\cdots,S_{l_3,2p_{l_3}})}},
\end{eqnarray}
where
$\rho_1=\dsum_{i=1}^{l_1}k_i,\rho_2=\dsum_{i=1}^{l_2}r_i,
\rho_3=\dsum_{i=1}^{l_3}(S_{i,1}+\cdots+S_{i,2p_i})$,
 and $\bar P_{(\sigma(k_1),\cdots,\sigma(k_{l_1}));(\sigma(r_1),\cdots,\sigma(r_{l_2}))}
            ^{(\sigma(S_{1,1}),\cdots;\cdots;
            \cdots,\sigma(S_{l_3,2p_{l_3}}))}$
is the coefficient of
$t_{k_1}\cdots t_{k_{l_1}}g_{r_1}\cdots g_{r_{l_2}}
t_{S_{1,1},\cdots,S_{1,2p_1}}\cdots
 t_{S_{l_3,1},\cdots,S_{l_3,2p_3}}$
in ${\bar W}^{m+1}$.

For example, we list some correlators
\begin{eqnarray}
\begin{array}{ll}
  \langle \tr M_1M_1^{\dag}\rangle=\langle \tr M_2M_2^{\dag}\rangle=\frac{1}{\mu} N^2,
  ~~~\langle \tr M_1M_1^{\dag}\tr M_2M_2^{\dag}\rangle=\frac{2}{\mu^2} N^4,\\
  \langle \tr M_1M_1^{\dag}\tr M_1M_1^{\dag}\rangle
    =\langle \tr M_2M_2^{\dag}\tr M_2M_2^{\dag}\rangle=\frac{1}{\mu^2}(N^2+1)N^2,
   \\
   \langle \tr M_1M_1^{\dag} M_2M_2^{\dag}\rangle=\frac{2}{\mu^2} N^3, ~~~
  \langle \tr (M_1M_1^{\dag})^3\rangle
    =\langle \tr (M_2M_2^{\dag})^3\rangle=\frac{6}{\mu^3}(N^2+N^4), \\
  \langle \tr M_1M_1^{\dag}\tr M_1M_1^{\dag}\tr M_1M_1^{\dag}\rangle
    =\langle \tr M_2M_2^{\dag}\tr M_2M_2^{\dag}\tr M_2M_2^{\dag}\rangle
    =\frac{1}{\mu^3}(N^2+2)(N^2+1)N^2, \\
  \langle \tr M_1M_1^{\dag}\tr M_1M_1^{\dag}\tr M_2M_2^{\dag}\rangle
    =\langle \tr M_1M_1^{\dag}\tr M_2M_2^{\dag}\tr M_2M_2^{\dag}\rangle
    =\frac{3}{\mu^3}(N^4+N^6), \\
  \langle \tr M_1M_1^{\dag}\tr M_1M_1^{\dag}M_2M_2^{\dag}\rangle
    =\langle \tr M_2M_2^{\dag}\tr M_1M_1^{\dag}M_2M_2^{\dag}\rangle
    =\frac{6}{\mu^3}(N^3+N^5),  \\
  \langle \tr M_1M_1^{\dag}\tr (M_2M_2^{\dag})^2\rangle
    =\langle \tr (M_1M_1^{\dag})^2\tr M_2M_2^{\dag}\rangle=\frac{8}{\mu^3} N^5, \\
  \langle \tr M_1M_1^{\dag}\tr (M_1M_1^{\dag})^2\rangle
    =\langle \tr M_2M_2^{\dag}\tr (M_2M_2^{\dag})^2\rangle=\frac{8}{\mu^3}(N^3+N^5).
\end{array}
\end{eqnarray}

\section{$W$-representation of fermionic two-matrix model}
The fermionic matrix model $Z_F$ with the super integrability is given by \cite{lywang2120}
\begin{eqnarray}\label{ZF}
Z_F&=&\frac{\int d\psi d\bar{\psi}
            \exp[N\sum_{k>0}\frac{p_k}{k}\tr(\bar{\psi} \psi)^k+N^2\tr(\bar{\psi} \psi)]}
{\int d\psi d\bar\psi\exp(N^2\tr\bar\psi\psi)}
\nonumber\\
&=&\sum_R(\frac{-1}{N})^{|R|}\frac{D_R(N)D_R(-N)}{d_R}S_R,
\end{eqnarray}
where $\psi$ and~$\bar{\psi}$ are independent complex Grassmann-valued~$N\times N$ matrices,
and $D_R(N) = S_R\{p_k = N\}$, $d_R = S_R\{p_k = \delta_{k,1}\}$ are respectively the dimension of
representation $R$ for the linear group $GL(N)$.

Let us extend (\ref{ZF}) to the fermionic two-matrix model,
\begin{eqnarray}\label{partition-F2}
Z_{2F}&=&\int d\psi d\bar{\psi}d\chi d\bar{\chi}
\exp[-\mu \tr \bar{\psi}\psi-\mu\tr\bar{\chi}\chi
     +\sum_{k=0}^{\infty}t_k \tr (\bar{\psi}\psi)^k
     +\sum_{k=0}^{\infty}g_k \tr ( \bar{\chi}\chi)^k
\nonumber\\&&
     +\sum_{l=1}^{\infty}\sum_{k_1,\cdots k_{2l}=1}^{\infty}t_{k_1,\cdots k_{2l}}
     \tr (\bar{\psi}\psi)^{k_1}(\bar{\chi}\chi)^{k_2}(\bar{\psi}\psi)^{k_3}\cdots
     (\bar{\psi}\psi)^{k_{2l-1}}(\bar{\chi}\chi)^{k_{2l}}],
\end{eqnarray}
where~$\chi$ and $\bar{\chi}$ are independent complex Grassmann-valued~$N\times N$ matrices.

There are the Virasoro constraints
\begin{eqnarray}
\check{\mathcal{L}}_nZ_{2F}=0,
\end{eqnarray}
where
\begin{eqnarray}\label{viraconst-2F}
\check{\mathcal{L}}_n &=&
\sum_{n=0}^{\infty}\sum_{s=1}^{n-1}\frac{\partial^2}{\partial t_{s}\partial t_{n-s}}
-\delta_{n,0}N^2
-\mu\sum_{n=0}^{\infty}\frac{\partial}{\partial t_{n+1}}
+\sum_{n=0}^{\infty}\sum_{k=0}^{\infty}kt_k\frac{\partial}{\partial t_{n+k}}
\nonumber\\&&
+\sum_{n=0}^{\infty}\sum_{l=1}^{\infty}\sum_{k_1,\cdots,k_{2l}=1}^{\infty}
 \sum_{a=1}^{l}k_{2a-1} t_{k_1,\cdots k_{2l}}
 \frac{\partial}{\partial t_{k_1,\cdots,n+k_{2a-1},k_{2a},\cdots,k_{2l}}}.
\end{eqnarray}

Similar to the complex two-matrix case, by considering the following infinitesimal transformations
in the integral (\ref{partition-F2}), respectively,

(i) $\psi\longrightarrow \psi+\epsilon\dsum_{n=0}^{\infty}(n+1)t_{n+1}\psi( \bar{\psi}\psi)^n$,

(ii)  $\psi\longrightarrow
      \psi+\epsilon\dsum_{n,m=0}^{\infty}[(n+1)+(m+1)]t_{n+1,m+1}
      \psi(\bar{\chi}\chi)^{m+1}( \bar{\psi}\psi)^{n}$,

(iii)  $\psi\longrightarrow
      \psi+\epsilon\dsum_{r=1}^{\infty}\dsum_{n_1,\cdots,n_{2r+1}=0}^{\infty}
      \check{\mathcal{N}}\check{T}
     \psi(\bar{\chi}\chi)^{n_2+1}( \bar{\psi}\psi)^{n_3+1}\cdots(\bar{\chi}\chi)^{n_{2r}+1}(\bar{\psi}\psi)^{n_{2r+1}}$,

(iv) $\chi\longrightarrow
      \chi+\epsilon\dsum_{m=0}^{\infty}(m+1)g_{m+1}\chi(\bar{\chi}\chi)^m$,

where $\check{\mathcal{N}}\check{T}=(n_2+1)+(n_3+1)+\cdots+(n_{2r+1}+1)t_{n_{2r+1},n_2+1,\cdots,n_{2r}+1}$,
we finally obtain
\begin{eqnarray}\label{constraint-F}
\mu\check{D}Z_{2F}=\check{W}Z_{2F},
\end{eqnarray}
where $\check{D}=\sum_{i=1}^4\check{D}_i$, $\check{W}=\sum_{i=1}^4\check{W}_i$,
the operators $\check{D}_i$ and $\check{W}_i$ are
\begin{eqnarray}\label{D-2F-1}
\begin{array}{ll}
\check{D_1}=
\dsum_{n=0}^{\infty}(n+1)t_{n+1}\frac{\partial}{\partial t_{n+1}},
\qquad\qquad~~~~
\check{D_2}=\dsum_{n,m=0}^{\infty}\check T_1\frac{\partial}{\partial
t_{n+1,m+1}},
\\
\check{D_3}=\dsum_{n_1,\cdots,n_{2r+1}=0}^{\infty}\check T_2
\frac{\partial}{\partial t_{n_1+1,\cdots,n_{2r+1}+1}},
\quad
\check{D_4}=\dsum_{m=0}^{\infty}(m+1)g_{m+1}\frac{\partial}{\partial g_{m+1}},
\end{array}
\end{eqnarray}

\begin{small}
\begin{eqnarray}\label{W-2F-1}
\check{W}_1&=&\sum_{n=0}^{\infty}(n+1)t_{n+1}
[
 \sum_{k=0}^{\infty}kt_{k}\frac{\partial}{\partial t_{n+k}}
 +\sum_{a=1}^{n-1}\frac{\partial}{\partial t_{a}}\frac{\partial}{\partial t_{n-a}}
 +\sum_{l=1}^{\infty}\sum_{a=1}^{l}\sum_{k_1,\cdots,k_{2l}=1}^{\infty}k_{2a-1}  \check T_0
 \nonumber\\&&
 \cdot
 \frac{\partial}{\partial t_{k_1,\cdots,k_{2a-1}+n,k_{2a},\cdots,k_{2l}}}]
-N^2t_{1},
\nonumber\\&&
\nonumber\\
\check{W_2}
&=&\sum_{n,m=0}^{\infty}\sum_{a=1}^{n-1}
\check T_1
\{\sum_{l=1}^{\infty}\sum_{k_1,\cdots,k_{2l}=1}^{\infty}
   \check T_0
    [\sum_{a=2}^{l}\sum_{s=0}^{k_{2a-1}}\frac{\partial}{\partial t_{k_1,\cdots,k_{2a-2},
     \check \xi_0, k_{2a},\cdots,k_{2l}}}
    +\sum_{s=0}^{k_{1}}\frac{\partial}{\partial t_{\check \xi_1,k_2, \cdots,k_{2l}}}]
\nonumber\\&&
+\frac{\partial}{\partial t_{a,m+1}}\frac{\partial}{\partial t_{n-a}}
  -N\delta_{n,0}\frac{\partial}{\partial g_{m+1}}%
 +\sum_{k=0}^{\infty}kt_{k}\frac{\partial}{\partial t_{n+k,m+1}}
\},
\nonumber\\&&
\nonumber\\
\check{W_3}
&=&\sum_{n_1,\cdots,n_{2r+1}=0}^{\infty}
\check T_2
\{[\sum_{s=1}^{n_{2r+1}-1}\frac{\partial}{\partial t_{s,n_2+1,\cdots,n_{2r}+1}}
   \frac{\partial}{\partial t_{n_{2r+1}-s}}
   +\sum_{b=1}^{r-1}\sum_{s=1}^{n_{2b+1}}\frac{\partial}{\partial t_{s,n_2+1,\cdots,n_{2b}+1}}
   \nonumber\\&&\cdot
   \frac{\partial}{\partial t_{\check \xi_2,n_{2b+2}+1,\cdots,n_{2r}+1}}
  ]
-N\delta_{n_{2r+1},0}\frac{\partial}{\partial t_{n_3+1,\cdots,n_{2r-1}+1,\check \xi_3}}
+\sum_{k=0}^{\infty}kt_k\frac{\partial}{\partial t_{n_{2r+1}+k,n_2+1,\cdots,n_{2r}+1}}
\nonumber\\&&
+\sum_{l=1}^{\infty}\sum_{k_1,\cdots,k_{2l}=1}^{\infty}
\check T_0
[\dsum_{s=0}^{k_1}
\frac{\partial}{\partial t_{s+1,n_2+1,\cdots,\check \xi_4,k_2,\cdots,k_{2l}}}
+\dsum_{a=2}^l\dsum_{s=0}^{k_{2a-1}}\frac{\partial}{\partial
t_{k_1,\cdots,k_{2a-2},s+1,\check \xi_5,k_{2a},\cdots,k_{2l}}}
]\},
\nonumber\\
&&
\nonumber\\
\check{W_4}
&=&\sum_{m=0}^{\infty}(m+1)g_{m+1}
[\sum_{a=1}^{m-1}\frac{\partial}{\partial g_{a}}\frac{\partial}{\partial g_{m-a}}
 +\sum_{k=0}^{\infty}kg_{k}\frac{\partial}{\partial g_{m+k}}
 +\sum_{l=1}^{\infty}\sum_{a=1}^{l}\sum_{k_1,\cdots,k_{2l}=1}^{\infty}k_{2a}
 \check T_0
 \nonumber\\&&\cdot
 \frac{\partial}{\partial t_{k_1,\cdots,k_{2a}+m,k_{2a+1},\cdots,k_{2l}}}
]
-N^2g_{1},
\end{eqnarray}
\end{small}
where
$\check T_0=t_{k_1,\cdots,k_{2l}}$,
$\check T_1=(n+m+2)t_{n+1,m+1}$,
$\check T_2=\bar{\mathcal{N}}t_{n_{2r+1}+1,n_2+1,\cdots,n_{2r}+1}$
and~$\check \xi_0=(a+1,m+1,n+k_{2a-1}-k-1)$,
~$\check \xi_1=(s+1,m+1,n+k_1-s-1)$,
~$\check \xi_2=n_{2r+1}+n_{2b+1}+1-s$,
~$\check \xi_3=n_{2r}+n_2+2$,
~$\check \xi_4=n_{2r+1}+k_1-1-s$,
~$\check \xi_5=(n_2+1,\cdots,n_{2r}+1,n_{2r+1}+k_{2a-1}-s-1)$.

We find that the fermionic two-matrix model (\ref{partition-F2})
can be realized by the $W$-representation
\begin{eqnarray}\label{wrep-F2}
Z_{2F}=e^{\frac{1}{\mu}\check{W}}e^{N(t_0+g_0)}.
\end{eqnarray}
The compact expression of correlators is
\begin{eqnarray}\label{correlator-F}
&&\langle\prod_{i=1}^{l_1} \tr (\bar{\psi}\psi)^{k_i}
         \prod_{j=1}^{l_2}\tr (\bar{\chi}\chi)^{r_j}
         \prod_{n=1}^{l_3} \tr(\bar{\psi}\psi)^{S_{n,1}}(\bar{\chi}\chi)^{S_{n,2}}\cdots
         (\bar{\psi}\psi)^{S_{n,2p_{n-1}}}(\bar{\chi}\chi)^{S_{n,2p_n}}
\rangle
\nonumber\\&&
=\frac{l_1!l_2!l_3!\dsum_{\rho_1+\rho_2+\rho_3=1}^{m+1}\sum_{\sigma}
       \check P_{(\sigma(k_1),\cdots,\sigma(k_{l_1}));(\sigma(r_1),\cdots,\sigma(r_{l_2}))}
               ^{(\sigma(S_{1,1}),\cdots,\sigma(S_{1,2p_1});\cdots;\sigma(S_{l_3,1}),\cdots,
                \sigma(S_{l_3,2p_{l_3}}))}}
{\mu^{m+1}(m+1)!\lambda_{(k_1,\cdots,k_{l_1})}\lambda_{(r_1,\cdots,r_{l_2})}
  \lambda_{(S_{1,1},\cdots,S_{1,2p_1};\cdots;S_{l_3,1},\cdots,S_{l_3,2p_{l_3}})}},
\end{eqnarray}
where
$\rho_1=\dsum_{i=1}^{l_1}k_i,\rho_2=\dsum_{i=1}^{l_2}r_i,
\rho_3=\dsum_{i=1}^{l_3}(S_{i,1}+\cdots+S_{i,2p_i})$,
and
$\check P_{(\sigma(k_1),\cdots,\sigma(k_{l_1}));(\sigma(r_1),\cdots,\sigma(r_{l_2}))}
               ^{(\sigma(S_{1,1}),\cdots;\cdots;\cdots,
                \sigma(S_{l_3,2p_{l_3}}))}$
is the coefficient of
$t_{k_1}\cdots t_{k_{l_1}}g_{r_1}\cdots g_{r_{l_2}}
t_{S_{1,1},\cdots,S_{1,2p_1}}
\cdots t_{S_{l_3,1},\cdots,S_{l_3,2p_3}}$ in ${\check W}^{m+1}$.

Here we list some correlators
\begin{eqnarray}
\begin{array}{ll}
 \langle \tr \bar{\psi}\psi\rangle=\langle \tr \bar{\chi}\chi\rangle=-\frac{1}{\mu} N^2,
 ~~\langle \tr \bar{\psi}\psi\tr \bar{\chi}\chi\rangle=\frac{2}{\mu^2} N^4,
  \\
 \langle \tr \bar{\psi}\psi\tr \bar{\psi}\psi\rangle
   =\langle \tr \bar{\chi}\chi\tr \bar{\chi}\chi\rangle=\frac{1}{\mu^2}(N^2-1)N^2,
   ~~ \langle \tr \bar{\psi}\psi \bar{\chi}\chi\rangle=-\frac{2}{\mu^2} N^3,
 \\
 \langle \tr (\bar{\psi}\psi)^3\rangle
   =\langle \tr (\bar{\chi}\chi)^3\rangle=\frac{6}{\mu^3}(-N^2+N^4),
  \\
  \langle \tr \bar{\psi}\psi\tr \bar{\psi}\psi\tr \bar{\psi}\psi\rangle
   =\langle \tr \bar{\chi}\chi\tr \bar{\chi}\chi\tr \bar{\chi}\chi\rangle
   =\frac{1}{\mu^3}(N^2+2)(N^2-1)N^2,\\
 \langle \tr \bar{\psi}\psi\tr \bar{\psi}\psi\tr \bar{\chi}\chi\rangle
   =\langle \tr \bar{\chi}\chi\tr \bar{\psi}\psi\tr \bar{\chi}\chi\rangle
   =\frac{3}{\mu^3}(N^4-N^6), \\
 \langle \tr \bar{\psi}\psi\tr \bar{\psi}\psi \bar{\chi}\chi\rangle
   =\langle \tr \bar{\chi}\chi\tr \bar{\psi}\psi \bar{\chi}\chi\rangle
   =-\frac{1}{\mu^3}(6N^3+2N^5)
 .
\end{array}
\end{eqnarray}

\section{Conclusion}
We have constructed the Hermitian, complex and fermionic two-matrix models with infinite set
of variables and presented their Virasoro constraints.
$W$-representation is important for understanding matrix model, since it provides a dual formula for
partition function through differentiation.
By considering the particular infinitesimal transformations of integration variables
in the partition functions, we finally derived
the desired operators preserving and increasing the grading. Thus it can be shown that
the two-matrix models constructed in this paper can be realized by the $W$-representations.
Moreover, by means of the $W$-representations, we derived the compact expressions of correlators
for these two-matrix models.
It should be noted that there are the infinite set of variables in these two-matrix models.
It leads to that we can not give their character expansions.
For further research, it would be interesting to study the case of $\beta$-deformed two-matrix models.

\setcounter{equation}{0}
\renewcommand\theequation{A.\arabic{equation}}
\begin{appendices}
\section{The operators $\hat W_i$ in (\ref{con-H1})}

\begin{small}
\begin{eqnarray}\label{W-H}
&&
\hat{W_1}=
\sum_{l=1}^{\infty}\sum_{k_1,\cdots ,k_{2l}=1}^{\infty}
\{t_1  {T_1}
[\delta_{k_1,1}
\frac{\partial}{\partial t_{k_3,\cdots,k_{2l-1},k_{2l}+k_{2}}}
+\sum_{a=2}^l\delta_{k_{2a-1},1}
\frac{\partial}{\partial t_{k_{1},\cdots,k_{2a-2}+k_{2a},\cdots,k_{2l}}
}
]
\nonumber\\
&&+\sum_{n=0}^{\infty}\sum_{a=1}^l
k_{2a-1}(n+1)t_{n+1}
\frac{\partial}{\partial t_{k_{1},\cdots,n+k_{2a-1}-1,\cdots,k_{2l}}
}
\}
+\sum_{k_{2}=1}^{\infty} t_1t_{1,k_2}\frac{\partial}{\partial g_{k_2}}
+t_1^2 N+2t_2N^2\nonumber\\
&&
+\sum_{n=0}^{\infty}(n+1)t_{n+1}
[\sum_{b=1}^{n-2}
\frac{\partial^2}{\partial t_b\partial t_{n-1-b}}
+\sum_{k=0}^{\infty}kt_k\frac{\partial}{\partial t_{n+k-1}}]
+\sum_{n=2}^{\infty}2N(n+1)t_{n+1}\frac{\partial}{\partial t_{n-1}}
,
\nonumber\\&&
\nonumber\\&&
\hat{W_2}=
\sum_{l,n=1}^{\infty}\sum_{k_{1},\cdots,k_{2l}=1}^{\infty}
  {T_2}
\{
(1-\delta_{k_1,1})
(\frac{\partial}{\partial t_{k_{1}-1,k_{2},\cdots,k_{2l}+n}}
+\frac{\partial}{\partial t_{k_{1}-1,k_2+n,\cdots,k_{2l}}})
\nonumber\\
&&+
\delta_{k_1,1}
\frac{\partial}{\partial
 t_{k_{3},\cdots,k_{2l}+n+k_{2}}
}
+\sum_{a=2}^l
(1-\delta_{k_{2a-1},1})
\frac{\partial}{\partial t_{k_{1},\cdots,k_{2a-2}+n,k_{2a-1}-1,k_{2a},\cdots
,k_{2l}}}
\nonumber\\
&&
+ \sum_{a=2}^{l-1}\delta_{k_{2a-1},1}
\frac{\partial}{\partial t_{k_{1},\cdots,k_{2a-2}+k_{2a}+n,\cdots ,k_{2l}}
}
\}
+
\sum_{\substack{l=2,\\n=1}}^{\infty}\sum_{k_{1},\cdots,k_{2l}=1}^{\infty}
 {T_2}
\delta_{k_{2l-1},1}
\frac{\partial}{\partial
 t_{k_{1},\cdots, k_{2l-2}+n+k_{2l}}
}
\nonumber\\
&&
+\sum_{l,n=1}^{\infty}\sum_{a=1}^{l}\sum_{k_{2a-1}=3}^{\infty}\sum_{b=1}^{k_{2a-1}-2}
\sum_{k_{1},\cdots, k_{2a-2}, \atop k_{2a},\cdots,k_{2l}=1}^{\infty}
 {T_2}[
(1-\delta_{a,1})
\frac{\partial}{\partial t_{k_{1},\cdots,k_{2a-2},b,n,k_{2a-1}-1-b,\cdots ,k_{2l}}}
\nonumber\\
&&
+\delta_{a,1}
\frac{\partial}{\partial
t_{b,n,k_{1}-1-b, \cdots ,k_{2l}}
}]
+\sum_{n=1}^{\infty}(n+1)t_{1,n}
[\sum_{k=2}^{\infty}kt_{k}\frac{\partial}{\partial t_{k-1,n}}
+\sum_{k_{2}=1}^{\infty}t_{1,k_{2}}\frac{\partial}{\partial g_{k_2+n}}
+t_1 \frac{\partial}{\partial g_{n}}],
\nonumber\\
&&
\nonumber\\
&&
\hat{W_3}
=\sum_{r=1}^{\infty}\sum_{n_1,\cdots,n_{2r}=1}^{\infty}
 {T_3}
\{N\delta_{n_1,1}
\frac{\partial}{\partial t_{n_3,\cdots,n_{2r}+n_2}}
+
(1-\delta_{n_1,1})
(\dsum_{a=1}^{n_{1}-2}\frac{\partial}{\partial t_{a}}
\frac{\partial}{\partial t_{n_{1}-1-a,\cdots,n_{2r}}}
\nonumber\\
&&
+
N\frac{\partial}{\partial t_{n_{1}-1,n_2,\cdots,n_{2r}}}
+\frac{\partial}{\partial t_{n_{1}-1}}
\frac{\partial}{\partial t_{n_3,\cdots,n_{2r}}}
)
+\sum_{s=2}^{r}\sum_{a=0}^{n_{2s-1}-2}(1-\delta_{n_{2s-1},1})
\frac{\partial}{\partial t_{n_1+a,n_2,\cdots,n_{2s-2}}}
\nonumber\\
&&\cdot
\frac{\partial}{\partial t_{n_{2s-1}-1-a,\cdots ,n_{2r}}}
+
\sum_{s=2}^{r-1}[\delta_{n_{2s-1},1}
\frac{\partial}{\partial t_{n_1,\cdots,n_{2s-2}}}
+
(1-\delta_{n_1,1})
\frac{\partial}{\partial t_{n_1+n_{2s-1}-1,n_2,\cdots,n_{2s-2}}}
\nonumber\\&&
\cdot
\frac{\partial}{\partial t_{n_{2s+1},\cdots,n_{2r}+n_{2s}}}
]
+(1-\delta_{n_{2r-1},1})
\frac{\partial}{\partial t_{n_1+n_{2r-1}-1,\cdots,n_{2r-2}}}
\frac{\partial}{\partial g_{n_{2r}}}
+
\sum_{k=0}^{\infty}kt_k \frac{\partial}{\partial
t_{n_{1}+k-1,\cdots,n_{2r}}}
\nonumber\\
&&
+\delta_{n_{2r-1},1}
\frac{\partial^2}{\partial t_{n_1,\cdots,n_{2r-2}}\partial g_{n_{2r}}}
+
\sum_{k_1,\cdots,k_{2l}=1}^{\infty}
 T_1
[
\sum_{i=1}^{l}
\sum_{s=0}^{k_{2i-1}-2}
\frac{\partial}{\partial t_{k_1,\cdots ,k_{2i-2},s+n_1,\cdots , n_{2r},
\xi_1,
\cdots,k_{2l}}}
\nonumber\\
&&
+\frac{\partial}{\partial t_{n_{1}+k_{1}-1,n_2\cdots ,n_{2r}+k_2,\cdots ,k_{2l}}}
+\frac{\partial}{\partial t_{k_1,\cdots,n_{1}+k_{2i-1}-1,\cdots,
n_{2r}+k_{2i},\cdots,k_{2l}}}
\}\nonumber\\
&&
+\dsum_{n_1=2}^{\infty}\dsum_{n_2=1}^{\infty}
(n_1+1+n_2)t_{n_1+1,n_2}
\frac{\partial^2}{\partial t_{n_1-1}\partial g_{n_2}}
,
\nonumber\\&&
\nonumber\\&&
\hat{W_4}=
\sum_{r=1}^{\infty}\sum_{n_2,\cdots,  n_{2r}=1}^{\infty}
 T_4
\{
\sum_{b=1}^{n_{3}-2}\frac{\partial^2}{\partial t_{b,n_2}\partial t_{n_{3}-1-b,\cdots ,n_{2r}}}
+(1-\delta_{n_3,1})(\frac{\partial^2}{\partial g_{n_2}\partial
t_{n_{3}-1,\cdots,n_{2r}}}
\nonumber\\
&&
+\frac{\partial^2}{\partial t_{n_{3}-1,n_2,}\partial t_{n_5,\cdots,n_{2r}+n_4}})
+\delta_{n_3,1}\frac{\partial^2}{\partial g_{n_2}\partial
t_{n_5,\cdots,n_{2r}+n_4}}
+\sum_{a=3}^{r-1}
[\delta_{n_{2a-1},1}
\frac{\partial}{\partial t_{n_3,\cdots ,n_{2a-2}+n_2}}
\nonumber\\
&&\cdot
\frac{\partial}{\partial t_{n_{2a+1},\cdots,n_{2r}}}
+(1-\delta_{n_{2a-1},1})(
\frac{\partial}{\partial t_{n_3,\cdots ,n_{2a-1}-1,n_{2a}}}
\frac{\partial}{\partial t_{n_{2a+1},\cdots,n_{2r}+n_{2a}}}
+
\frac{\partial}{\partial t_{n_3,\cdots ,n_{2a-2}+n_2}}
\nonumber\\
&&
\cdot\frac{\partial}{\partial t_{n_{2a-1}-1,n_{2a}\cdots,n_{2r}}})
]
+[\delta_{n_{2r-1},1}\frac{\partial}{\partial t_{n_3,\cdots ,n_{2r-2}+n_2}}
+(1-\delta_{n_{2r-1},1})
\frac{\partial}{\partial t_{n_3,\cdots , n_{2r-1}-1,n_2}}
]\frac{\partial}{\partial g_{n_{2r}}}
\nonumber\\
&&
+\sum_{a=3}^{r}\sum_{b=1}^{n_{2a-1}-2}
\frac{\partial}{\partial t_{n_3,\cdots,n_{2a-2},b,n_2}}
\frac{\partial}{\partial t_{n_{2a-1}-1-b,\cdots ,n_{2r}}}
+t_1\frac{\partial}{\partial t_{n_3,\cdots ,n_{2r}+n_2}}
+\sum_{k=2}^{\infty}kt_k
\frac{\partial}{\partial t_{k-1,n_2,\cdots ,n_{2r}}}
\nonumber\\
&&
+\sum_{l=1}^{\infty}\sum_{k_1,\cdots,k_{2l}=1}^{\infty}
  T_1
[\delta_{k_1,1}
\frac{\partial}{\partial t_{n_3,\cdots,n_{2r}+k_2,\cdots ,k_{2l}+n_2}}
+(1-\delta_{k_1,1})
(\frac{\partial}{\partial t_{k_1-1,n_2,\cdots ,n_{2r}+k_2,\cdots ,k_{2l}}}
\nonumber\\
&&
+
\frac{\partial}{\partial t_{n_3,\cdots,n_{2r},k_1-1,\cdots
,k_{2l}+n_2}})
+
\sum_{b=2}^{l}
(1-\delta_{k_{2b-1},1})
(\frac{\partial}{\partial t_{k_1,\cdots,k_{2b-3},\xi_2,\cdots,k_{2l}}}
+\frac{\partial}{\partial t_{k_1,\cdots,\xi_3,\cdots, n_{2r}+k_{2l}}}
)
\nonumber\\&&
+\sum_{b=2}^{l}
\delta_{k_{2b-1},1}
\frac{\partial}{\partial t_{k_1,\cdots
,k_{2b-1}+n_2,\xi_4,\cdots,k_{2l}}}
]
+
\sum_{l=1}^{\infty}\sum_{a=1}^{l}\sum_{k_{2a-1}=3}^{\infty}
\sum_{k_1,\cdots,k_{2a-2},\atop k_{2a},\cdots,k_{2l}=1}^{\infty}
\sum_{s=1}^{k_{2a-1}-2}
  T_1
\frac{\partial}{\partial t_{k_1,\cdots,k_{2a-2},s,\xi_5,\cdots,k_{2l}}}
\},
\nonumber\\
&&
\nonumber\\
&&
\hat{W_5}=g_1^2 N
+\sum_{n,k=0}^{\infty}(n+1)g_{n+1}kg_k\frac{\partial}{\partial g_{n+k-1}}
+\sum_{k_l,k_2,k_3=1}^{\infty}g_1t_{k_1,k_2,k_3,1}\frac{\partial}{\partial
t_{k_1+k_3,k_2} }
+\sum_{k_1=1}^{\infty}t_{k_1,1}g_1\frac{\partial}{\partial t_{k_1}}
\nonumber\\
&&
+\sum_{l= 1}^{\infty}\sum_{n= 0}^{\infty}
\sum_{k_1,\cdots,k_{2l}=1}^{\infty}
  T_5
[
\delta_{k_{2l},1}\sum_{a=1}^{l-1}
\frac{\partial}{\partial t_{k_1+k_{2l-1},\cdots,k_{2l-2}}}
k_{2a}\frac{\partial}{\partial t_{k_1,\cdots,
\xi_6,\cdots,k_{2l}}}
+
\frac{\partial}{\partial t_{k_1+k_{2l-1},\cdots,k_{2l-2}}}
\cdot
\nonumber\\
&&
\cdot
\frac{\partial}{\partial t_{k_1,\cdots ,
k_{2l-1},\xi_7}}
+
\delta_{n,0}(1- \delta_{k_{2a},1})
\frac{\partial}{\partial
t_{k_{1},k_{2}-1,\cdots,k_{2l}}}
+\delta_{n,0}
\sum_{a=1}^{l-1}
 \delta_{k_{2a},1}\frac{\partial}{\partial t_{k_1,\cdots
, k_{2a-1}+k_{2a+1},\cdots ,k_{2l}}}
]
\nonumber\\&&
+\sum_{n=1}^{\infty}\sum_{s=1}^{n-2}(n+1)g_{n+1}\frac{\partial}{\partial
g_s} \frac{\partial}{\partial g_{n-1-s} }
+
\sum_{n=2}^{\infty}2N(n+1)g_{n+1}\frac{\partial}{\partial
g_{n-1}}
+2g_{2}N^2,
\end{eqnarray}\end{small}
where
$T_1=t_{k_1,\cdots,k_{2l}}$,
~$  T_2=(n+1)t_{1,n}t_{k_{1},\cdots,k_{2l}}$,
~$ T_3=\mathcal{N}_1t_{n_1+1,n_2,\cdots,n_{2r}}$,
~$  T_4=\mathcal{N}_2t_{1,n_2,\cdots,n_{2r}}$,
~$ T_5=(n+1)g_{n+1}t_{k_1,\cdots k_{2l}}$ and
 $\xi_1=k_{2i-1}-1-s$,
 $\xi_2=(k_{2b-2}+n_2,\cdots, n_{2r},k_{2b-1}-1)$,
 $\xi_3=(k_{2b-1}-1,n_2)$,
 $\xi_4=(n_3,\cdots,n_{2r}+k_{2b})$,
 $\xi_5=(n_2,\cdots,n_{2r},k_{2a-1}-1-s)$,
 $\xi_6=k_{2a}+n-1$,
 $\xi_7 =k_{2l}+n-1$.
\end{appendices}

\section *{Acknowledgment}

This work is supported by the National Natural Science Foundation
of China (No. 11875194).

\end{document}